\documentclass{lipics-v2021}

\makeatletter
\renewenvironment{abstract}{%
  \vskip\topmattervskip\bigskipamount
  \noindent
  \rlap{\color{lipicsLineGray}\vrule\@width\textwidth\@height1\p@}%
  \hspace*{7mm}\fboxsep1.5mm\colorbox[rgb]{1,1,1}{\raisebox{-0.4ex}{%
    \large\selectfont\sffamily\bfseries\abstractname}}%
  \vskip3\p@
  \fontsize{9}{12}\selectfont
  \noindent\ignorespaces}
  {%
  \protected@write\@auxout{}{\string\gdef\string\@pageNumberEndAbstract{\thepage}}%
  }
\makeatother

\author{Nils Lommen}{RWTH Aachen University, Aachen, Germany}{lommen@cs.rwth-aachen.de}{https://orcid.org/0000-0003-3187-9217}{}
\author{Moritz Leven Rosarius}{RWTH Aachen University, Aachen, Germany}{lommen@cs.rwth-aachen.de}{https://orcid.org/0009-0009-4419-4525}{}
\author{Jürgen Giesl}{RWTH Aachen University, Aachen, Germany}{lommen@cs.rwth-aachen.de}{https://orcid.org/0000-0003-0283-8520}{}
\authorrunning{Nils Lommen, Moritz Leven Rosarius, and Jürgen Giesl}

\usepackage{mathtools,amsfonts,amsmath,amssymb}
\usepackage{tikz}
\usetikzlibrary{shapes,calc,arrows,automata,decorations.pathmorphing,decorations.pathreplacing,positioning}

\usepackage{hyperref}
\hypersetup{
  pdftitle={ Verifying Parameterized LTL for Infinite State Systems via Termination Analysis}, colorlinks=true, linkcolor=blue, citecolor=olive, filecolor=magenta, urlcolor=cyan
}

\usepackage[capitalize,nameinlink]{cleveref}
\usepackage{todonotes}
\setuptodonotes{inline}

\usepackage{multicol}
\usepackage{hyperref}
\usepackage{tabto}
\usepackage{thm-restate}
\usepackage{stmaryrd}
\usepackage{booktabs}

\usepackage[shortlabels,inline]{enumitem}
\setlist[enumerate,1]{label=(\alph*), wide=0pt, leftmargin=*}
\setlist[itemize,1]{label=\textbullet}

\makeatletter
\newcommand{\inlineitem}[1][]{%
  \ifnum\enit@type=\tw@
    {\descriptionlabel{#1}}
    \hspace{\labelsep}%
  \else \ifnum\enit@type=\z@ \refstepcounter{\@listctr}\fi \quad\@itemlabel\hspace{\labelsep}%
  \fi}

\newcommand{\tool}[1]{\textsf{#1}}

\newcommand{\braced}[1]{\left\lbrace #1 \right\rbrace}
\newcommand{\var}{\normalfont\texttt}
\newcommand{\wildcard}{\underline{\hspace{0.15cm}}}

\newcommand{\eval}[2]{\llbracket #1 \rrbracket_{#2}}

\newcommand{\NN}{\mathbb{N}}
\newcommand{\ZZ}{\mathbb{Z}}

\newcommand{\identity}{{\normalfont\textsf{id}}}
\newcommand{\abs}[1]{\lvert #1 \rvert}

\providecommand{\monus}{% Don't redefine it if available
  \mathbin{% We want a binary operation
    \vphantom{+}% The same height as a plus or minus
    \text{% Change size in sub/superscripts
      \mathsurround=0pt % To be on the safe side
      \ooalign{% Superimpose the two symbols
        \noalign{\kern-.35ex}% but the dot is raised a bit
        \hidewidth$\smash{\cdot}$\hidewidth\cr % Dot
        \noalign{\kern.35ex}% Backup for vertical alignment
        $-$\cr % Minus
      }%
    }%
  }%
}

\newcommand{\AtomSet}{\mathcal{A}}

\newcommand{\true}{\var{true}}
\newcommand{\false}{\var{false}}

\newcommand{\valuation}{\sigma}
\newcommand{\Valuation}{\Sigma}

\newcommand{\VSet}{\mathcal{V}}

\newcommand{\guard}{g}
\newcommand{\update}{\eta}
\newcommand{\Program}{\mathcal{P}}
\newcommand{\TSet}{\mathcal{T}}
\newcommand{\LSet}{\mathcal{L}}

\newcommand{\location}{\ell}
\newcommand{\locationInitial}{\tilde{\location}}
\newcommand{\locationTerminal}{\location_\bot}
\newcommand{\ITS}{(\VSet,\LSet,\locationInitial,\locationTerminal,\TSet)}

\newcommand{\trueLTL}{\true}
\newcommand{\falseLTL}{\false}
\newcommand{\nextLTL}{\mathbf{X}\,}
\newcommand{\until}{\,\mathbf{U}\,}

\newcommand{\globally}{\mathbf{G}\,}

\newcommand{\eventually}{\mathbf{F}\,}

\newcommand{\at}{\mathsf{at}}
\newcommand{\FairStates}{F}

\crefname{definition}{Def.}{Def.}
\crefname{example}{Ex.}{Ex.}
\crefname{counterexample}{Counterex.}{Counterex.}
\crefname{appendix}{App.}{App.}
\crefname{ex}{Ex.}{Ex.}
\crefname{theorem}{Thm.}{Thm.}
\crefname{thm}{Thm.}{Thm.}
\crefname{lemma}{Lemma}{Lemmas}
\crefname{lem}{Lemma}{Lemmas}
\crefname{remark}{Rem.}{Rem.}
\crefname{obs}{Obs.}{Obs.}
\crefname{section}{Sect.}{Sect.}
\crefname{subsection}{Sect.}{Sect.}
\crefname{subsubsection}{Sect.}{Sect.}
\crefname{line}{Line}{Lines}
\crefname{corollary}{Cor.}{Cor.}
\crefname{cor}{Cor.}{Cor.}
\crefname{figure}{Fig.}{Fig.}
\crefname{enumi}{}{}
\crefname{algorithm}{Alg.}{Alg.}

\nolinenumbers
\hideLIPIcs

\title{Verifying LTL for Infinite State Systems via Termination Analysis}

\allowdisplaybreaks

\begin{document}
\maketitle
\begin{abstract}
  We show that existing tools for termination analysis are extremely well suited for LTL model checking of infinite state systems.
  To this end, we present a framework \tool{MoAT} which uses the well-known automata-based approach and reduces the LTL model checking problem to fair termination.
  To prove or disprove fair termination, it then calls the termination tools \tool{KoAT} and \tool{LoAT} in the backend.
  Our experiments show that in this way, \tool{MoAT} is on par with existing state-of-the-art tools for LTL model checking of infinite state systems.
\end{abstract}

\section{Introduction}
Linear temporal logic (LTL) is widely used to express properties of software.
In particular, LTL formulas can be used to specify local or global invariants of programs and they can also express properties like termination.
There is a wealth of work on deciding LTL properties for finite state systems.
However, the abstraction step from general programs to finite state systems naturally results in a loss of precision.
Therefore, there is a growing interest in verifying LTL properties for infinite state systems (see, e.g., \cite{brockschmidtT2TemporalProperty2016,chatterjee2025SoundCompleteWitnesses,cimatti2025InfiniteStateLivenessChecking,dietsch2015FairnessModuloTheory,unnoModularPrimalDualFixpoint2023}).
To tackle this problem, we use a framework which constructs the synchronized product of a Büchi automaton capturing the analyzed LTL formula and an integer transition system (ITS) modeling the program.
We apply our tools \tool{KoAT} \cite{JAR26} and \tool{LoAT} \cite{frohn2022ProvingNonTerminationLower} for termination analysis to check \emph{fair termination} of the product ITS, which solves the original verification problem.
We implemented our approach in a new tool \tool{MoAT} (Model Checking Analysis Tool) and demonstrate its strengths via an experimental evaluation, which compares \tool{MoAT} to other state-of-the-art tools.
While fair termination is undecidable in general, \tool{MoAT} benefits from integrating \tool{KoAT} and \tool{LoAT}, which can often prove or refute this property for a wide range of practical programs.
Our evaluation demonstrates that \tool{MoAT} is on par with the leading tools for LTL model checking of infinite state systems.
Thus, this shows that state-of-the-art tools for termination analysis are also very suitable for applications in other areas.

We define ITSs in \Cref{sect:ITS} and LTL in \Cref{sect:LTL_syntax}.
Then we recapitulate the reduction of LTL verification to fair termination in \Cref{sec:LTL-fair_termination}.
Finally, in \Cref{sect:evaluation} we discuss how the existing termination tools \tool{KoAT} and \tool{LoAT} can be used to check fair termination and we provide an experimental evaluation to show their strength when using them for LTL model checking.

\section{Integer Transition Systems}
\label{sect:ITS}
ITSs are a widely studied formalism in program verification (see, e.g., \cite{abdulla1996GeneralDecidabilityTheorems,brockschmidt2016AnalyzingRuntimeSize,chatterjee2025SoundCompleteWitnesses,dingel1995ModelCheckingInfinite,frohn2022ProvingNonTerminationLower,henzinger1995ComputingSimulationsFinite,ruemmer2018Eldarica,kupferman2000AutomataTheoreticApproachReasoning}).
An ITS is a tuple $\ITS$ with a finite set of variables $\VSet$, a finite set of locations $\LSet$, an initial location $\locationInitial \in \LSet$, a terminal location $\locationTerminal$, and a finite set of transitions $\TSet$.
A \emph{transition} is a 4-tuple $(\location,\guard,\update,\location')$ with a \emph{start location} $\location\in\LSet$, \emph{target location} $\location'\in\LSet\setminus\{\locationInitial\}$, and a polynomial \emph{update} function $\update: \VSet\rightarrow\ZZ[\VSet]$.
We require that transitions with start location $\locationTerminal$ also have target location $\locationTerminal$.
In addition, each transition has a guard $\guard$, which is a conjunction of polynomial inequations of the form $p_1 < p_2$ where $p_1, p_2 \in \ZZ[\VSet]$.
As syntactic sugar, we also allow inequations like $p_1 \geq p_2$, $\true$, or $\false$.
We use the terminal location $\locationTerminal$ to extend finite ``stuck'' evaluations to infinite ones.
This will allow us to perform model checking via fair termination of the synchronized product of the ITS with a suitable Büchi automaton.
\begin{example}
  In the ITS of \cref{fig:ITS}, we omitted identity updates $\update(v) = v$ and guards where $\guard$ is $\true$.
  Here, we have  $\VSet = \braced{x,y}$ and  $\LSet = \{\location_0, \location_1, \location_2, \locationTerminal \}$ where $\location_0$ is initial.
   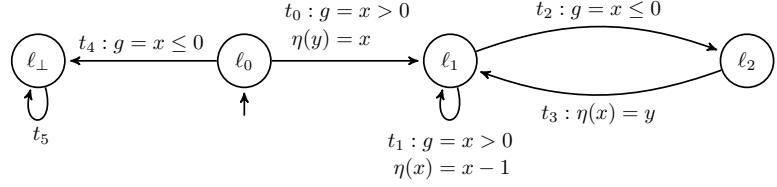
\begin{figure}[t]
    \begin{center}
      \scalebox{0.8}{
        \begin{tikzpicture}[->,>=stealth',shorten >=2pt,auto,node distance=4cm,thick,initial text=$ $]
          \node[state,initial below] (l0) {$\location_0$};
          \node[state] (l1) [right = 2.5cm of l0]{$\location_1$};
          \node[state] (lT) [left = 2.5cm of l0]{$\locationTerminal$};
          \node[state] (l2) [right = of l1]{$\location_2$};
          \draw (l0) edge [above] node [align=center] {
              $
                \begin{array}{r@{\;}c@{\;}l}
                  t_0: \guard & = & x > 0 \\
                  \update(y)  & = & x
                \end{array}
              $} (l1);
          \draw (l1) edge [bend left = 20] node [align=center,above] {$t_2: \guard = x \leq 0$} (l2);
          \draw (l2) edge [bend left = 20] node [align=center,below] {$t_3: \update(x) = y$} (l1);
          \draw (l1) edge [loop below] node [] {
              $
                \begin{array}{r@{\;}c@{\;}l}
                  t_1: \guard & = & x > 0 \\
                  \update(x)  & = & x - 1
                \end{array}
              $
            } (l1);
          \draw (l0) edge node [align=center,above] {$t_4: \guard = x \leq 0$} (lT);
          \draw (lT) edge [align=center,loop below] node [xshift=0.05cm] {$t_5$} (lT);
        \end{tikzpicture}
      }
    \end{center}
    \vspace*{-.5cm}
    \caption{An Integer Transition System}
    \label{fig:ITS}
    \vspace*{-.3cm}
  \end{figure}
\end{example}

From now on, we fix an ITS $\ITS$.
To define the semantics of ITSs, the values of the variables are ``stored'' in a \emph{valuation} $\valuation: \VSet \rightarrow \ZZ$, where $\Valuation$ denotes the set of all valuations.
When evaluating a transition $(\location, \guard, \update, \location')$, the values of the variables are changed according to its update $\update$.
So an evaluation step moves from one \emph{configuration} $(\location,\valuation)\in\LSet\times\Valuation$ to another configuration $(\location',\valuation')$  where $\valuation\models\guard$ holds.
Here, the valuation $\valuation'$ is obtained by applying the update $\update$ on $\valuation$.
Let $\eval{e}{\valuation}$ denote the \emph{evaluation} of an expression $e$ in the valuation $\valuation\in\Valuation$, where every variable $v$ in $e$ is replaced by its value $\valuation(v)$.
So evaluating $\eval{3\cdot x}{\valuation}$ and $\eval{x > 0}{\valuation}$ at $\valuation(x) = 2$ results in $6$ and $\true$, respectively.
\begin{definition}[Evaluation of ITSs]
  \label{def:EvaluationITS}
  For two configurations $(\location,\valuation)$ and $(\location',\valuation')$, and a transition $t = (\location_t,\guard,\update,\location_{t}')\in\TSet$, $(\location,\valuation)\rightarrow_t(\location',\valuation')$ is an \emph{evaluation step} by $t$ if
  \begin{itemize}
    \TabPositions{6.2cm}
    \itemsep0em
    \item $\location = \location_t$ and $\location' = \location_{t}'$, \tab \inlineitem[] $\valuation\models\guard$, and
    \item for every program variable $v\in\VSet$ we have $\valuation'(v) = \eval{\update(v)}{\valuation}$.
  \end{itemize}
  The union of all relations $\to_t$ for $t \in \TSet$ is denoted by $\to_{\TSet}$.
  We write $\to$ instead of $\to_t$ or $\to_{\TSet}$, if $t$ or $\TSet$ are clear from the context.
  We abbreviate an \emph{evaluation} $(\location_0,\valuation_0)\rightarrow \cdots \rightarrow(\location_k,\valuation_k)$ by $(\location_0,\valuation_0)\rightarrow^k(\location_k,\valuation_k)$.
  A \emph{run} is an infinite evaluation.
  An evaluation or a run is called \emph{initial} if it starts in the initial location $\locationInitial$.
  Note that despite finiteness of $\VSet$, $\LSet$, and $\TSet$, an ITS usually has infinitely many evaluations.
\end{definition}
\begin{example}
  \label{ex:evaluation}
  Let us denote valuations $\valuation\in\Valuation$ as tuples $(x,y) \in \ZZ^2$.
  Then the ITS in \cref{fig:ITS} has the evaluation $(\location_0,(2,0)) \to_{t_0} (\location_1,(2,2)) \to_{t_1}^2 (\location_1,(0,2)) \to_{t_2} (\location_2,(0,2)) \to_{t_3} (\location_1,(2,2))$.
  Note that the evaluation $(\location_1,(2,2)) \to^4 (\location_1,(2,2))$ can be repeated infinitely often.
  Thus, the ITS in \Cref{fig:ITS} is non-terminating.
  Properties like termination, safety, or reachability can easily be expressed as LTL formulas.
\end{example}

\section{LTL Formulas}
\label{sect:LTL_syntax}
An \emph{LTL formula} is of the form
\[
  \varphi \coloneq p_1 < p_2 \mid \at(\location) \mid \neg \varphi \mid \varphi \land \varphi \mid \varphi \lor \varphi \mid \nextLTL \varphi \mid \varphi \until \varphi
\]
where $p_1, p_2 \in \ZZ[\VSet]$ are polynomials and $\location\in\LSet$.
Here, $\at(\location)$ holds if a run starts in $\location$.
Furthermore, we use other Boolean connectives and operators like ``eventually'' $\eventually \varphi = \trueLTL \until \varphi$ and ``globally'' $\globally \varphi = \neg \eventually \neg \varphi$, etc.\ as syntactic sugar.
We recapitulate the standard semantics of LTL (see, e.g., \cite{Huth_Ryan_2004}).
Let $\pi = (\location_0, \valuation_0) \to (\location_1,\valuation_1)\to \dots$ be a run.
For any $i \in \NN$, let $\pi^{\geq i}$ denote the suffix $(\location_i,\valuation_i)\to (\location_{i + 1},\valuation_{i + 1}) \to \dots$ of $\pi$.
Let $[k-1]$ denote the set $\braced{0,\dots,k-1}$.
Then we have
\begin{itemize}
  \TabPositions{6.2cm}
  \itemsep0em
  \item $\pi \models p_1 < p_2$ iff $\valuation_0 \models p_1 < p_2$ \tab \inlineitem $\pi \models \at(\location)$ iff $\location_0 = \location$,
  \item $\pi \models (\varphi \land \psi)$ iff $\pi \models \varphi$ and $\pi \models \psi$, \tab \inlineitem $\pi \models (\varphi \lor \psi)$ iff $\pi \models \varphi$ or $\pi \models \psi$,
  \item $\pi \models \neg\varphi$ iff $\pi \not\models \varphi$, \tab \inlineitem $\pi \models \nextLTL\varphi$ iff $\pi^{\geq 1} \models \varphi$,
  \item $\pi \models \varphi\until\psi$ iff there exists $k\in\NN$ with $\pi^{\geq i}\models \varphi$ for all $i \in [k-1]$ and $\pi^{\geq k} \models \psi$.
\end{itemize}
For an LTL formula $\varphi$, we call $L(\varphi) = \braced{\pi\mid\pi\models\varphi}$ the \emph{language} of $\varphi$.
\begin{example}
  \label{ex:LTL_formulas}
  Consider the LTL formulas $\eventually \at(\location_\bot)$ and $\globally (\at(\location_1) \rightarrow \eventually (x \leq 0))$.
  The first formula states that the terminal location is eventually reached, i.e., the program terminates.
  The second formula expresses that whenever one is in location $\location_1$, then one eventually reaches a configuration where $x \leq 0$ holds.
\end{example}

\begin{definition}[Verification and Refutation of LTL Formulas]
  Let $\varphi$ be an LTL formula.
  Then
  \begin{itemize}
    \item $\varphi$ is called \emph{verified} iff $\pi \models\varphi$ holds for all initial runs $\pi$ and
    \item $\varphi$ is called \emph{refuted} iff $\pi \models\neg\varphi$ holds for an initial run $\pi$.
  \end{itemize}
\end{definition}

Note that both $\varphi$ and $\neg \varphi$ can be refuted (if both formulas are satisfiable).
So in particular, refuting $\neg \varphi$ does not verify $\varphi$.
\begin{example}
  \label{ex:LTLformulasVerified}
  The ITS of \cref{fig:ITS} is non-terminating, i.e., $\eventually \at(\location_\bot)$ is refuted, see Ex.\ \ref{ex:evaluation}.
  In \Cref{sec:LTL-fair_termination} we will verify the LTL formula $\globally (\at(\location_1) \rightarrow \eventually (x \leq 0))$ for this ITS.
\end{example}

\section{Verifying LTL via Fair Termination}
\label{sec:LTL-fair_termination}
We now recapitulate the classical reduction of LTL verification to \emph{fair termination} based on Büchi automata.
For fair termination, only a designated subset of locations or transitions is required to be evaluated finitely often.
In \Cref{sect:evaluation}, we will discuss how this can be automatically checked by the tools \tool{KoAT} and \tool{LoAT} in the framework \tool{MoAT}.
In particular, we will also provide an experimental evaluation demonstrating the strengths of \tool{MoAT}.

\begin{figure}[t]
\begin{center}
  \scalebox{0.8}{
    \begin{tikzpicture}[->,>=stealth',shorten >=2pt,auto,node distance=5cm,thick,initial text=$ $]
      \node[state,initial left] (l0) {$q_0$};
      \node[state,accepting] (l1) [right = 3cm of l0]{$q_1$};
      \draw (l0) edge [align=center,loop above] node {$e_1: \true$} (l0);
      \draw (l0) edge node [align=center,above] {$e_2: \at(\location_1) \land x>0$} (l1);
      \draw (l1) edge [align=center,loop right] node {$e_3: x > 0$} (l1);
    \end{tikzpicture}
  }
\end{center}
\vspace*{-.3cm}
\caption{Büchi Automaton for the LTL Formula $\varphi: \; \eventually (\at(\location_1) \land \globally (x > 0))$}
\label{fig:BA}
\vspace*{-.3cm}
\end{figure}

A (non-deterministic) \emph{Büchi automaton} \cite{richardbuchi1966SymposiumDecisionProblems} is a tuple $\mathcal{B} = (Q,A,\delta,\tilde{q},F)$ where $Q$ is a finite set of states, $A$ is a finite alphabet, $\delta \subseteq (Q \times A) \times Q$ is the transition relation, $\tilde{q} \in Q$ is the initial state, and $F\subseteq Q$ is the set of accepting (or ``fair'') states.
The Büchi automaton $\mathcal{B}$ \emph{accepts} the infinite word $w = w_0 w_1 \dots$ with $w_i \in A$ iff there is a run of $\mathcal{B}$ with $q_i \stackrel{w_i}{\longrightarrow} q_{i + 1}$ (i.e., $(q_i,w_i,q_{i + 1}) \in \delta$) for all $i \in \NN$ where $q_0 = \tilde{q}$ and a state from $F$ is visited infinitely often.
$L(\mathcal{B})$ is the \emph{language} of $\mathcal{B}$, i.e., the set of all accepted words.
For every LTL formula $\varphi$ one can compute a Büchi automaton $\mathcal{B}$ such that $L(\varphi)$ \emph{corresponds} to $L(\mathcal{B})$ \cite{Vardi_1996}.
\begin{example}
  Consider the Büchi automaton in \Cref{fig:BA}, where double circles denote fair states.
  Its alphabet are all subsets of $\varphi$'s atoms $\AtomSet_\varphi = \{ \at(\location_1), x > 0 \}$ and the edges in \Cref{fig:BA} represent several transitions.
  For example, the edge $e_3$ represents the two transitions $(q_1, \{x > 0\}, q_1)$ and $(q_1, \{x > 0, \at(\location_1)\}, q_1)$, while the edge $e_1$ represents four transitions from $q_0$ to $q_0$ (for the four subsets of $\AtomSet_\varphi$).
  In general, an edge $(q, \guard, q')$ represents all transitions $(q, M, q')$ with $M \subseteq \AtomSet_\varphi$ where $(\bigwedge_{p \in M} p) \land (\bigwedge_{p \in \AtomSet_\varphi \setminus M} \lnot p)$ implies $\guard$.
  The language accepted by the automaton in \Cref{fig:BA} corresponds to the language of the LTL formula $\varphi = \eventually (\at(\location_1) \land \globally (x > 0))$.
\end{example}

LTL verification can be reduced to fair termination.
We call an ITS  \emph{fairly terminating} on a set $\LSet_f\subseteq\LSet$ iff there is no initial run that visits some location of $\LSet_f$ infinitely often.
We call such a run fairly terminating.
\begin{definition}[Fair Termination]
  A run $(\location_0,\valuation_0) \to (\location_1,\valuation_1) \to \dots$ is \emph{fairly terminating} on the set $\LSet_f \subseteq \LSet$ if $\abs{\braced{i\in\NN\mid\location_i\in \LSet_f}} < \infty$.
  An ITS is \emph{fairly terminating} on $\LSet_f$ if every initial run is fairly terminating on $\LSet_f$.
\end{definition}

Now, we recapitulate the reduction to fair termination (see, e.g., \cite{vardi1994ReasoningInfiniteComputations}).
Let $\Program = \ITS$ be an ITS, $\varphi$ an LTL formula, and $\mathcal{B}_{\neg\varphi} = (Q,2^{\AtomSet_{\varphi}},\delta,\tilde{q},F)$ a Büchi automaton whose language corresponds to $L(\neg\varphi)$.
We define the \emph{synchronized product} $\Program\otimes\mathcal{B}_{\neg\varphi}$ as the ITS $(\VSet,(\LSet\times Q) \cup \{ \locationTerminal \}, (\locationInitial,\tilde{q}),\locationTerminal,\TSet')$.
Here, $\TSet'$ consists of all transitions $((\location,q),\guard\land \guard_M,\update,(\location',q'))$ whenever there is a transition $(\location,\guard,\update,\location')\in\TSet$ and $(q,M,q') \in \delta$ for some $q,q'\in Q$ and $M \subseteq \AtomSet_{\varphi}$, where $\guard_M$ results from $\bigwedge_{p \in M} p \land \bigwedge_{p \in \AtomSet_{\varphi} \setminus M} \neg p$ by replacing the atom $\at(\location)$ by $\trueLTL$ and all atoms $\at(\hat{\location})$ with $\hat{\location} \neq \location$ by $\falseLTL$.
Moreover, for every location $(\location,q)\in\LSet\times Q$, $\TSet'$ must contain a transition $((\location,q), \neg\guard_1 \land \ldots \land \neg \guard_n,\identity,\locationTerminal)$, where $\guard_1,\dots,\guard_n$ are the guards of the remaining outgoing transitions of $(\location,q)$.
The following theorem recapitulates soundness and completeness of this reduction in our setting.
\begin{restatable}[From Büchi Automata to Fair Termination]{theorem}{ltlConstruction}
  \label{lem:ltlConstruction}
  Let $\Program$ be an ITS and $\mathcal{B}_{\neg\varphi}$ be a Büchi automaton such that $L(\mathcal{B}_{\neg\varphi})$ corresponds to $L(\neg\varphi)$.
  Then the ITS $\Program\otimes\mathcal{B}_{\neg\varphi}$ is fairly terminating on $\LSet\times\FairStates$ iff $\varphi$ is verified.
\end{restatable}

\begin{example}
  \label{ex:refutation}
  \Cref{fig:Crossproduct} shows the ITS $\Program\otimes\mathcal{B}_{\neg\varphi}$ for the ITS $\Program$ from \Cref{fig:ITS} and the Büchi automaton $\mathcal{B}_{\neg\varphi}$ from \Cref{fig:BA}.
  Here, we omitted transitions with unsatisfiable guards and unreachable locations.
  The only fair location is $(\location_1,q_1)$, depicted by a double circle.
  Thus, $\Program\otimes\mathcal{B}_{\neg\varphi}$ is fairly terminating as the only relevant cycle $\{t_7\}$ cannot be executed infinitely often.
  Hence, $\globally (\at(\location_1) \rightarrow \eventually (x \leq 0))$ is verified.
  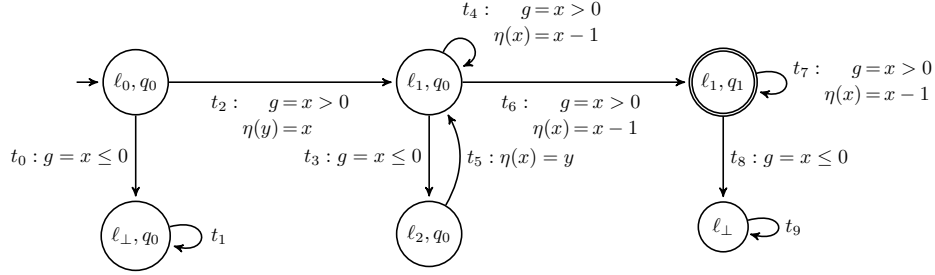
\begin{figure}[t]
    \scalebox{0.75}{
      \begin{tikzpicture}[->,>=stealth',shorten >=2pt,auto,node distance=5cm,thick,initial text=$ $]
        \node[state,initial left] (l0) {$\location_0,q_0$};
        \node[state] (lT) [below = 1.5cm of l0] {$\locationTerminal,q_0$};
        \node[state] (l1) [right = 4cm of l0] {$\location_1,q_0$};
        \node[state] (l2) [below = 1.5cm of l1] {$\location_2,q_0$};
        \node[state,accepting] (l1a) [right = 4cm of l1] {$\location_1,q_1$};
        \node[state] (lTT) [below = 1.5cm of l1a] {$\location_\bot$};
        \draw (l0) edge node [align=center,left] {$t_0: \guard = x \leq 0$} (lT);
        \draw (lT) edge [align=center,loop right, looseness=6] node [xshift=0.05cm] {$t_1$} (lT);
        \draw (l0) edge [align=center,below] node [xshift=0.05cm] {$t_2: \hspace*{-.2cm}
              \begin{array}[t]{r@{\,}c@{\,}l}
                \guard     & = & x > 0 \\
                \update(y) & = & x
              \end{array}
            $} (l1);
        \draw (l1) edge node [align=center,left] {$t_3: \guard = x \leq 0$} (l2);
        \draw (l1) edge [loop, out=65, in=25, looseness=4] node [xshift=-0.25cm, yshift=-0.25cm] {$t_4: \hspace*{-.2cm}
              \begin{array}[t]{r@{\,}c@{\,}l}
                \guard     & = & x > 0 \\
                \update(x) & = & x - 1
              \end{array}
            $} (l1);
        \draw (l2) edge [bend right] node [align=center,right] {$t_5: \update(x) = y$} (l1);
        \draw (l1) edge node [align=center,below] {$t_6: \hspace*{-.2cm}
              \begin{array}[t]{r@{\,}c@{\,}l}
                \guard     & = & x > 0 \\
                \update(x) & = & x-1
              \end{array}
            $} (l1a);
        \draw (l1a) edge [loop right, looseness=6] node {$t_{7}: \hspace*{-.2cm}
              \begin{array}[t]{r@{\,}c@{\,}l}
                \guard     & = & x > 0 \\
                \update(x) & = & x - 1
              \end{array}
            $} (l1a);
        \draw (l1a) edge node [right]
          {$t_{8}: \guard = x \leq 0$}(lTT);
        \draw (lTT) edge [loop right] node
          {$t_{9}$} (lTT);
      \end{tikzpicture}
    }
    \caption{An Integer Transition System Representing $\Program\otimes\mathcal{B}_{\neg\varphi}$}
    \label{fig:Crossproduct}
    \vspace*{-.4cm}
  \end{figure}
\end{example}

\section{Implementation and Evaluation}
\label{sect:evaluation}

\begin{table}[t]
  \begin{center}
    \scalebox{0.7}{
    \begin{tabular}
      {
        l @{\hspace{1em}} ccc p{0.5pt} ccc p{0.5pt} ccc p{0.5pt} ccc p{0.5pt} ccc
      }
      \toprule Tool    & \multicolumn{3}{c}{\tool{MoAT}} &     & \multicolumn{3}{c}{\tool{MuVal}} &  & \multicolumn{3}{c}{\tool{nuXmv}} &     & \multicolumn{3}{c}{\tool{Ultimate}} &  & \multicolumn{3}{c}{\tool{LTL-VerP}}                                                             \\
      \cmidrule(lr){2-4} \cmidrule(lr){6-8} \cmidrule(lr){10-12} \cmidrule(lr){14-16} \cmidrule(lr){18-20}
      Prop.\           & Yes                             & No  & Total                            &  & Yes                              & No  & Total                               &  & Yes                                 & No  & Total &  & Yes & No  & Total &  & Yes & No  & Total \\
      RA               & 65                              & 239 & 304                              &  & 44                               & 241 & 285                                 &  & 5                                   & 221 & 226   &  & 54  & 164 & 218   &  & 9   & 191 & 200   \\
      OV               & 16                              & 295 & 311                              &  & 15                               & 298 & 313                                 &  & 7                                   & 297 & 304   &  & 16  & 270 & 286   &  & 11  & 35  & 46    \\
      RC               & 107                             & 198 & 305                              &  & 108                              & 199 & 307                                 &  & 101                                 & 193 & 294   &  & 98  & 153 & 251   &  & 89  & 41  & 130   \\
      PR               & 103                             & 199 & 302                              &  & 106                              & 201 & 307                                 &  & 104                                 & 193 & 297   &  & 99  & 153 & 252   &  & 82  & 54  & 136   \\
      \midrule Total   & \multicolumn{3}{c}{1222}        &     & \multicolumn{3}{c}{1212}         &  & \multicolumn{3}{c}{1121}         &     & \multicolumn{3}{c}{1007}            &  & \multicolumn{3}{c}{512}                                                                         \\
      \midrule Avg.\ T & \multicolumn{3}{c}{1.11s}       &     & \multicolumn{3}{c}{5.02s}        &  & \multicolumn{3}{c}{7.26s}        &     & \multicolumn{3}{c}{8.90s}           &  & \multicolumn{3}{c}{33.81s}                                                                      \\
      Avg.+
      T                & \multicolumn{3}{c}{0.78s}       &     & \multicolumn{3}{c}{2.52s}        &  & \multicolumn{3}{c}{0.80s}        &     & \multicolumn{3}{c}{3.07s}           &  & \multicolumn{3}{c}{4.76s}                                                                       \\
      \bottomrule
    \end{tabular}
    }
  \end{center}
  \caption{Evaluation on LTL Formulas}
  \label{table:evaluation}
  \vspace*{-.5cm}
\end{table}

We now discuss our implementation in the tool \tool{MoAT} and provide an experimental evaluation which compares \tool{MoAT} with other state-of-the-art tools for LTL verification of infinite state systems.
To obtain Büchi automata for classical LTL formulas, \tool{MoAT} makes use of existing libraries like \tool{Spot}~\cite{lutz2022Spot} which generate optimized (smaller) \emph{transition-based accepting} Büchi automata whose acceptance condition relies on visiting a fair transition (instead of a fair state) infinitely often.
The approach via fair termination works in the same way for such Büchi automata.

To prove and disprove fair termination, \tool{MoAT} uses the tools \tool{KoAT} \cite{JAR26} and \tool{LoAT} \cite{frohn2022ProvingNonTerminationLower}.
However, up to now \tool{KoAT} could only prove and \tool{LoAT} could only disprove \emph{ordinary} termination of ITSs.
Therefore, we now explain how they can be adapted to analyze \emph{fair} termination.
\subsection{Proving Fair Termination}
An ITS is fairly terminating on a location if and only if none of its incoming transitions can be executed infinitely often.
\tool{KoAT} uses a \emph{modular} approach to prove ordinary termination.
More precisely, \tool{KoAT} infers a termination argument for each transition individually.
Thus, to prove \emph{fair termination} one only has to consider the result of \tool{KoAT} on all incoming transitions of fair locations.
If \tool{KoAT} proves termination for all these transitions, then the ITS is fairly terminating.
This is captured in the following observation.
\begin{observation}[Fair Termination via Incoming Transitions]
  \label{lem:fair_term_incoming}
  An ITS is fairly terminating on a set of fair locations $\FairStates$ if and only if all transitions $(\wildcard,\wildcard,\wildcard,\location)$ with $\location\in\FairStates$ are terminating.
\end{observation}

So for example, one could prove fair termination by assigning a ranking function to each location such that the values of the ranking functions are bounded and decreasing on transitions to fair locations and non-increasing on all other transitions.
This ensures that there is no run which reaches fair locations infinitely often.
To infer termination arguments, \tool{KoAT} uses several sophisticated improvements of this technique, e.g., it implements multiphase-linear ranking functions \cite{BenAmramGenaimCAV17,giesl2022ImprovingAutomaticComplexity} and it uses a modular approach such that it only has to consider sub-programs when computing ranking functions.
Furthermore, \tool{KoAT} uses a specific technique to prove termination for triangular weakly non-linear loops \cite{FMSD25,JAR26} and control-flow refinement.

\subsection{Disproving Fair Termination}
\tool{LoAT} proves non-termination by searching for reachable non-terminating cycles via an acceleration driven clause learning calculus \cite{frohn2022ProvingNonTerminationLower,frohn2023ProvingNonTerminationAcceleration}.
This calculus is suitable for arbitrary forms of ITSs, including ITSs with nested loops.
A cycle is \emph{fairly} non-terminating if and only if one of its locations is a fair location.
Thus, to disprove fair termination, one just has to run \tool{LoAT} until it finds a reachable non-terminating cycle that contains a fair location.
The counterexample trace found by \tool{LoAT} is a trace for the product ITS.
Hence its states are pairs of ITS locations and states of the automaton.
To obtain a counterexample trace on the level of the original ITS, one has to project each of these pairs to the respective original ITS location.

\subsection{Experiments}
We compare our implementation to state-of-the-art tools for verifying LTL properties for infinite state systems.
More precisely, we compare \tool{MoAT} with \tool{LTL-VerP} \cite{chatterjee2025SoundCompleteWitnesses}, \tool{Ultimate} \cite{dietsch2015FairnessModuloTheory}, \tool{MuVal} \cite{unnoModularPrimalDualFixpoint2023}, and \tool{nuXmv} \cite{cimatti2014nuXmv}.
For each tool that supports several configurations, we selected the configuration recommended by the authors.
For \tool{LTL-VerP}, the results reported in \cite{chatterjee2025SoundCompleteWitnesses} were obtained by evaluating a large number of parameter configurations and selecting the best-performing one for each benchmark instance.
In contrast, our experiments apply the single configuration identified by the authors as the overall best setting.
This configuration uses linear templates for ranking functions.
Consequently, the  performance of \tool{LTL-VerP} in our experiments is not directly comparable to the one in \cite{chatterjee2025SoundCompleteWitnesses}.
The tool \tool{nuXmv} also has several different configurations.
For our experiments, we used the \tool{IC3} engines.

We considered the same benchmarks and similar LTL properties as in \cite{chatterjee2025SoundCompleteWitnesses}.
This benchmark set consists of 318 \tool{C} programs from the \emph{Software Verification Competition} (\emph{SV-COMP} \cite{SV-Comp}) and the \emph{Termination Competition} (\emph{TermComp} \cite{termcomp}).
To run \tool{MoAT} on these programs, we translated them into ITSs.
We used the following LTL formulas, which we adapted slightly compared to \cite{chatterjee2025SoundCompleteWitnesses}:
\begin{itemize}
  \TabPositions{3.1cm}
  \itemsep0em
  \item Reach-Avoid (RA): \tab $(\eventually \at(\locationTerminal)) \land (\globally (x \geq 0))$
  \item Overflow (OV): \tab $\eventually(x < -64 \lor x > 63)$
  \item Recurrence (RC): \tab $\globally\eventually (x \geq 0)$
  \item Progress (PR): \tab $\globally(x < -5 \rightarrow \eventually (x > 0))$
\end{itemize}
Here, ``$x$'' refers to the variable which is the first when sorting the variable names alphabetically.
The LTL formula RA expresses that the terminal location is reached and a negative value of $x$ is avoided.
OV holds if $x$ overflows, RC holds if a non-negative value of $x$ is reached infinitely often, and PR holds if a value of $x$ less than $-5$ eventually becomes positive.
The results of our evaluation are depicted in \Cref{table:evaluation} where we used a timeout of $60$s per example as in \emph{TermComp}.
All tools were run inside an Ubuntu Docker container on a machine with an AMD Ryzen 7 3700X octa-core CPU and $8 \, \mathrm{GB}$ of RAM.
As mentioned, in \tool{MoAT}'s backend, \tool{KoAT} was used for the ``Yes'' answers and \tool{LoAT} was used for answering~``No''.

In our experiments, while \tool{MuVal} was slightly better than \tool{MoAT} in some categories, in total \tool{MoAT} solved the most benchmarks and achieved near-complete coverage ($>96\%$) across all categories.
``Avg.+ T'' denotes the average runtime of successful runs in seconds, whereas ``Avg.\ T'' is the average runtime of all runs.
\tool{MoAT} was the fastest tool with an average runtime of $1.11$s per example.

\bibliographystyle{plainurl}
\bibliography{bib}

\end{document}